\documentclass[aps,notitlepage,twocolumn,nofootinbib,pra,10pt]{revtex4-2}
\usepackage{amsmath,amsfonts,amssymb,amsthm,bbm,graphicx,enumerate,times,mathtools,hyperref,physics,marvosym,wasysym}
\usepackage[ruled,longend]{algorithm2e}
\usepackage[dvipsnames]{xcolor}
\usepackage{soul}
\usepackage{url}

\makeatletter 
\renewcommand{\thefigure}{S\@arabic\c@figure}
\makeatother
\begin{document}

\title{Conformal Properties of Hyperinvariant Tensor Networks}
\author{Matthew Steinberg$^{1,2*}$}
\author{Javier Prior$^{3,4}$}
\affiliation{$^{1}$QuTech, Delft University of Technology, Delft, The Netherlands}
\affiliation{$^{2}$Quantum and Computer Engineering Department, Delft University of Technology, Delft, The Netherlands}
\affiliation{$^{3}$ Departamento de F\'isica Aplicada, Universidad Polit\'ecnica de Cartagena, Cartagena 30202, Spain}
\affiliation{$^{4}$Instituto Carlos I de F\'isica Te\'orica y Computacional, Universidad de Granada, Granada 18071, Spain}

\date{\today}

\begin{abstract}
Hyperinvariant tensor networks (hyMERA) were introduced as a way to combine the successes of \textit{perfect tensor networks} (HaPPY) and the \textit{multiscale entanglement renormalization ansatz} (MERA) in simulations of the AdS/CFT correspondence. Although this new class of tensor network shows much potential for simulating conformal field theories arising from hyperbolic bulk manifolds with quasiperiodic boundaries, many issues are unresolved. In this manuscript we analyze the challenges related to optimizing tensors in a hyMERA with respect to some quasiperiodic critical spin chain, and compare with standard approaches in MERA. Additionally, we show two new sets of tensor decompositions which exhibit different properties from the original construction, implying that the multitensor constraints are neither unique, nor difficult to find, and that a generalization of the analytical tensor forms used up until now may exist. Lastly, we perform randomized trials using a descending superoperator with several of the investigated tensor decompositions, and find that the constraints imposed on the spectra of local descending superoperators in hyMERA are compatible with the operator spectra of several minimial model CFTs.
\end{abstract}

\maketitle


\section{Introduction}\label{introduction}

Tensor networks have proven to be excellent theoretical tools for investigating different aspects of many-body quantum entanglement at low energies for local Hamiltonians. Due to their success, tensor networks have found applications in many different fields, some of which may not be obvious from the original motivation of tensor-network theory \cite{orus1,orus2,javi-bh,TNs_ML}. Although many different models have emerged over the years (with such notable examples as MPS, PEPS, and TTN) proving to be very useful for the parameterization of local, gapped Hamiltonians \cite{biamonte1,bridgeman1}), tensor networks such as the \textit{multiscale entanglement renormalization ansatz} (or MERA) have also proven to be useful in the simulation of quantum critical lattice models \cite{PRL-QuantumMERAchannels}; such work also concerns the study of conformal field theory (or CFT) \cite{vidal1,vidal2,evenbly1,evenbly2,evenbly3,eisert_unify,PRL-QuantumMERAchannels} and the AdS/CFT correspondence \cite{maldacena1,witten,klebanov}. 

Although many of the details of these tensor networks are quite distinct, the fundamentals remain the same: a tensor network is an efficient parameterization of certain classes of states, determined in turn by the geometry evoked in the network, which takes the form of a diagrammatic representation of the entanglement degrees of freedom for a quantum state. The actual network consists of a series of tensors that are related via interconnected contractions and open indices. The open indices of the tensor network are attributed to the physical degrees of freedom of the system in question. Contractions between these tensors in a tensor-network geometry help to define the particular auxiliary degrees of freedom of a quantum system in the form of quantum entanglement. It is using this method that very large Hilbert spaces can be broken down into parameterizations for only very specific parts of this large Hilbert space. Such quantum states generally obey an area law, in which the entanglement entropy of a subsystem scales with its associated boundary area, and not like its volume. However, it is well-known that quantum critical systems violate the area law of entanglement entropy \cite{orus1,area_laws} by way of introducing a logarithmic correction term; the MERA has been used to great effect in this area of active theoretical research \cite{vidal1,vidal2,evenbly1,evenbly2,evenbly3,orus2,evenbly_disordered,PRL-QuantumMERAchannels}. A brief review of MERA is given in Appendix \ref{MERA_family}, with examples shown in Figure \ref{MERA_family_TNs}.     

It was first noted in \cite{swingle1,swingle2} that MERA exhibits some similarities with the AdS/CFT correspondence \cite{maldacena1,witten,klebanov}, as it was already used successfully for extracting conformal data from a quantum critical lattice system. However, it has been shown by the work of \cite{beny1,carroll1,qCFT,czech_kinematic,czech_defect} that MERA exhibits preferred directionality due to requirements imposed on isometric and unitary tensors in the tensor network; these arguments rely on the fact that the tensor network can be viewed as a toy-model analog to a hyperbolic-bulk discretization of AdS space. Preferred directionality is generally not a feature of uniform bulk AdS space. Additionally, other research works have suggested that tensor-network analogs that exhibit a holographic duality can be better studied in systems with impurities or disorder \cite{CQCs,czech_defect,qCFT,evenbly_minimal}. 

A few years later, following the work of Harlow \textit{et al.} \cite{harlow_bulk}, a new proposal for a toy model of holography was published \cite{happy}, as shown in Figure \ref{tilings} of Appendix \ref{perfect_tensors}. \textit{Perfect tensor networks} were developed in part due to observations that bulk-local operators act on particular subspaces of the boundary CFT, much in the same way that a quantum error correction code utilizes local-logical operators to project and encode onto subspaces of a quantum system. Perfect tensor networks realize many aspects of the AdS/CFT correspondence, such as an exact, discretized version of the Ryu-Takayanagi formula \cite{takayanagi,takayanagi1,Ryu_2006}. However, perfect tensor networks inscribed on periodic boundaries become problematic when calculating $n$-point correlation functions. Due in part to the ease of describing stabilizer states in a perfect tensor network, any $2$-point correlation function will result in a trivial phase. This behavior is not typically seen in a CFT-groundstate, which generally exhibits long-range quantum entanglement. The existence of algebraically decaying correlation functions \cite{infinitehappy} are a necesary condition for simulation of CFT-groundstates. Jahn \textit{et al.} found that quasiperiodic boundaries of regular hyperbolic tilings can, on average, give rise to correlation function decays and average entanglement entropy scaling that are similar to those observed in CFTs. As a result of their efforts, perfect tensor networks can be associated to the so-called \textit{Majorana dimer states} in the bulk, giving rise to strongly-disordered critical spin chains on the boundary \cite{majorana_dimers,central_dimers,qCFT,jahn2021holographictopical}.

In spite of the successes of both of the MERA and perfect tensor networks, one may ask whether a class of tensor-network ans{\"a}tze exists which generalizes the desired properties of both models nontrivially. An answer was given with \textit{hyperinvariant tensor networks} (or hyMERA) \cite{hyper_evenbly}, which proposed relaxed constraints on the bulk tensors of MERA; such conditions come in the form of the \textit{multitensor constraints}. hyMERA networks possess an interesting array of features in their own right: algebraically decaying correlation functions; nontrivial entanglement spectra; bulk uniform-symmetry characteristics; analytical descriptions of constituent bulk isometries and unitaries; and the preservation of local, real-space renormalization-group steps via unitary and isometric tensors (as in MERA). Although a hyMERA network seems to capture many of the novel characteristics of a tessellated slice of AdS spacetime in principle, there are still many technical questions to be answered that impede its utilization as a practical tensor-network ansatz for simulating CFTs in the setting of holography. 

In this paper, we address some of these questions. Firstly, we discuss why any hyMERA tensor network is not currently optimizable using a local-Hamiltonian variational groundstate optimization, and discuss possible resolutions to this issue. Secondly, we introduce several new tensors (which we name $S$ and $T$) that do not exhibit the same properties of the original tensors from \cite{hyper_evenbly}; indeed, these tensors are not even unitary. Using these new tensors, we show that many different possible solutions to the multitensor constraints can be found using only the original tensor-decomposition structure. These findings indicate that solutions to the multitensor constraints are not unique, and that a tensor generalization of the strict analytical forms for $Y,Q,R,S$ and $T$ may exist. Lastly, we perform a randomized simulation of a descending superoperator used in hyMERA with several of these tensor decompositions, and show that the resulting bounds on scaling dimensions encapsulate several minimal model CFTs. Our work shows that if both a suitable optimization scheme and tensor generalization can be developed, then hyMERA can be employed as a numerical tool for investigating CFTs arising from quasiperiodic critical spin chains.

The structure of the article is as follows:  Section \ref{hyper_explain}  provides an introduction to the structure of hyperinvariant tensor networks and the originally-proposed tensor decompositions. Section \ref{challenges_hyMERA}  explains the difficulties associated with quasiperiodicity on the hyMERA boundary. In Section \ref{results_section} we propose two new example tensor decompositions, and provide numerical evidence that the multitensor constraints allow for the scaling dimensions of several minimal model CFTs to be reproduced. In Section \ref{discussion-section}, we discuss these results and propose future research directions.

\section{Hyperinvariant Tensor Networks}\label{hyper_explain}

\textit{Hyperinvariant tensor networks} (or hyMERA) have been proposed as a way to combine the most desired characteristics of MERA and perfect tensor networks, with the goal of modeling the symmetry and entanglement characteristics of a timeslice of AdS spacetime in a discretized, toy-model setting.

\begin{figure}
\centering
\includegraphics[width=\columnwidth]{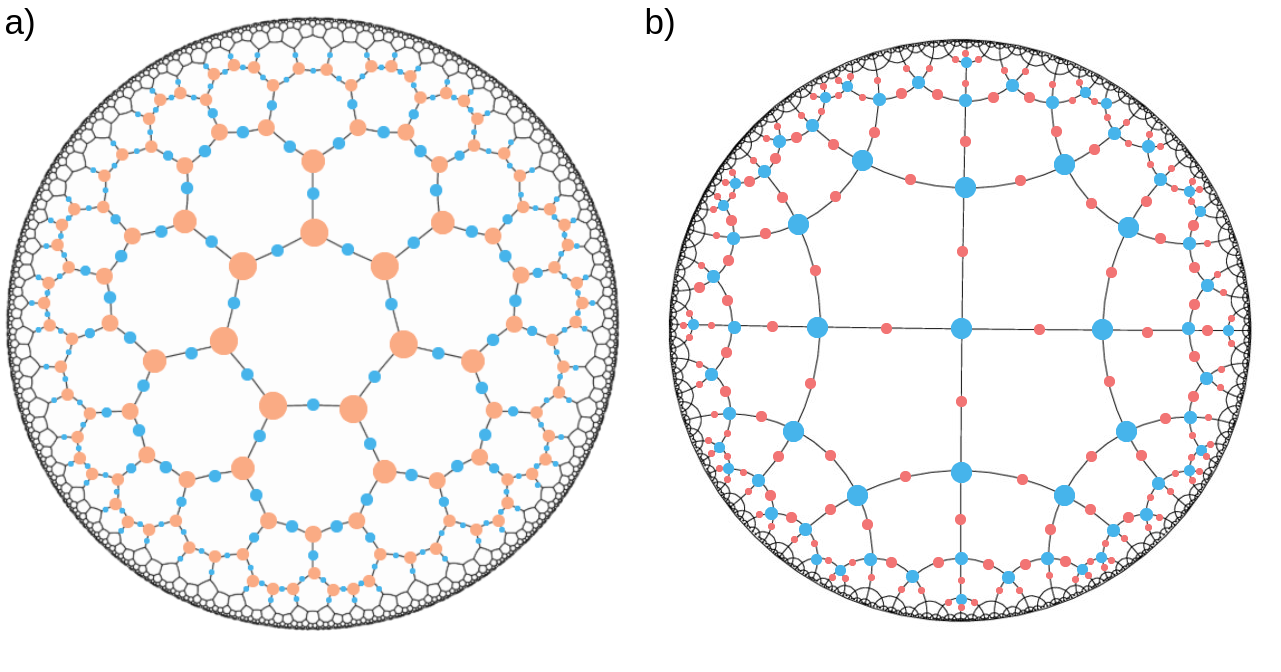}
\caption{The $\{5,4\}$ and $\{7,3\}$ tilings of a hyMERA network, respectively. Every hyMERA network is comprised of a Schl{\"a}fli number $\{p,q\}$ tiling that describes the geometric relationship of the bulk to its boundary.}
\label{hyMERA_tilings}
\end{figure}

A hyMERA network is constructed from a hyperbolic tessellation; examples of these tessellations can be seen in Figure \ref{hyMERA_tilings} . Each hyperbolic tessellation is defined by a Schl{\"a}fli number $\{p,q\}$ in the bulk manifold ($p$ represents the number of edges per polygon in the bulk manifold, and $q$ is the number of edges meeting at each vertex), with the requirement that $\frac{1}{p} + \frac{1}{q} < \frac{1}{2}$ for hyperbolic manifolds. Each of these $\{p,q\}$-structured manifolds we will call \textit{parameterizations} of hyMERA. The network can be arranged into concentric layers of tensors with which one can perform entanglement renormalization. Every concentric layer $\mathcal{L}$ of the hyMERA network's bulk can be described as realizing a step in the scale-invariant, realspace RG-flow of a typical MERA network; in the thermodynamic limit, the scale factors $s$ of the network differ significantly from MERA, depending on the parameterization employed. For example, $\{7,3\}$-hyMERA and $\{5,4\}$-hyMERA both exhibit irrational scale factors $s = 2.\overline{618}$ and $3.7\overline{32}$, respectively. 

\begin{figure}
\centering
\includegraphics[width=\columnwidth]{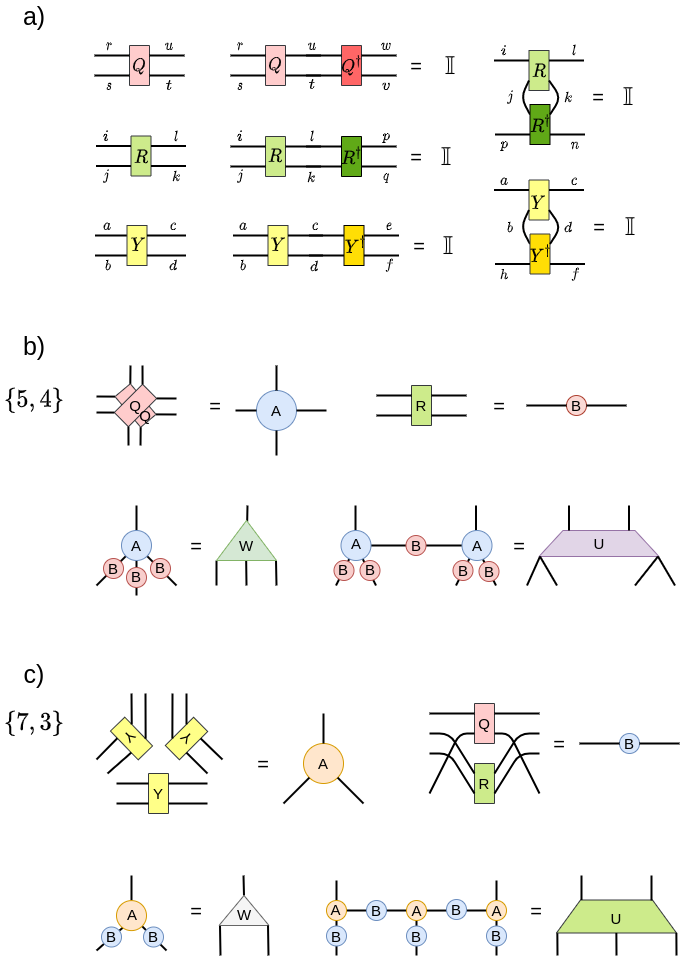}
\caption{Subfigure $a)$ shows the example particular tensor decomposition solution applied to the set $\{A,B\}$ for each of the two parameterizations of hyMERA considered in this article. The requisite relations that the set $\{Y,Q,R\}$ for multitensor-constraint adherence in a hyMERA network are displayed; We additionally require that the finite set of reduced-density matrices $\rho(\mathcal{R}_{n})$ must also exhibit nontrivial entanglement spectra. Subfigure $b)$ shows the construction of the sets of tensors $\{A,B\}$ and $\{U,W\}$ from consituent tensors $\{Y,Q,R\}$ in the $\{7,3\}$ parameterization. Finally, Subfigure $c)$ displays the construction of the sets of tensors $\{A,B\}$ and $\{U,W\}$ from consituent tensors $\{Q,R\}$ in the $\{5,4\}$ version.} 
\label{multilinear_hyMERA}
\end{figure}

Figure \ref{multilinear_hyMERA} describes how to build a parameterization of hyMERA, using a set of example unitaries and isometries $\{U(A,B),W(A,B)\}$, which are needed in order to realize entanglement renormalization on the $\{p,q\}$-manifold.

Every vertex and edge in a hyMERA nework consists of a set of tensors which we shall call $\{A,B\}$, where $A$ corresponds to tensors with $q$ indices, and $B$ represents rank-$2$ tensors on every edge of the tessellation. How to decompose the set $\{A,B\}$ depends on the particular \textit{tensor decomposition} that is chosen; at present, the best-known tensor decomposition has been introduced in \cite{hyper_evenbly}. For our work, we implemented the same tensor decomposition in the $\{7,3\}$ and $\{5,4\}$ versions of hyMERA, which are described in detail in Figure \ref{multilinear_hyMERA}. In the solution provided, tensors $\{A,B\}$ are constructed from a set of smaller tensors $\{Y,Q,R\}$ for the $\{7,3\}$ parameterization and from the set $\{Q,R\}$ for the $\{5,4\}$-parameterization. These constituent tensors take the explicit forms 
\begin{align}
Y_{abcd} &=
\begin{bmatrix}
\cos\theta_{1} & 0 & 0 & i\sin \theta_{1} \\
0 & \sin\theta_{1} & i\cos \theta_{1} & 0 \\
0 & i\cos \theta_{1} & \sin \theta_{1} & 0 \\
i\sin \theta_{1} & 0 & 0 & \cos \theta_{1} \\
\end{bmatrix}, \\
R_{efgh} &=
\begin{bmatrix}
\cos \theta_{2} & 0 & 0 & i\sin \theta_{2} \\
0 & \cos \theta_{2} & i\sin \theta_{2} & 0 \\
0 & i\sin \theta_{2} & \cos \theta_{2} & 0 \\
i\sin \theta_{2} & 0 & 0 & \cos \theta_{2} \\
\end{bmatrix}, \\
Q_{ijkl} &=
\begin{bmatrix}
\cos \theta_{3} & 0 & 0 & (\sin \theta_{3})e^{i\theta{4}} \\
0 & \cos \theta_{5} & i\sin \theta_{5} & 0 \\
0 & i\sin \theta_{5} & \cos \theta_{5} & 0 \\
(\sin\theta_{3})e^{i\theta{4}} & 0 & 0 & -(\cos \theta_{3})e^{2i\theta{4}} \\
\end{bmatrix}.
\end{align}
The set of tensors $\{Y,Q,R\}$ are governed by five free parameters $\{\theta_{i}\}_{i=1}^{5}$; moreover, the bond dimension of each hyMERA parameterization can be broken down into smaller constituent indices, each with a bond dimension that recombines in such a way so as to recover the original bond dimension seen in the tessellation \cite{hyper_evenbly}. The solution to the multitensor constraints that we used takes this form due to the fact that the set $\{Y,Q,R\}$ must adhere to what are known as the \textit{multitensor constraints}, which are intended to provide more symmetry in the bulk of a hyMERA network, but at the same time to be compatible with notions of entanglement renormalization. Every solution to the multitensor constraints must satisfy certain general properties: 
\begin{enumerate}
\item The multitensor constraints must be parameterized by some set of so-called free-parameters $\{\theta_{i}\}_{i=1}^{n}$. 
\item These free parameters $\{\theta_{i}\}_{i=1}^{n}$ \textit{must increase} as the bond dimension $\chi$ of the network increases.
\item Finally, networks must be generated which have nontrivial entanglement spectra and correlation functions. More succinctly, the entanglement spectra must depend on the free parameters, and for some $n$-site reduced-density matrix, the spectra \textit{must not} be proportional to the $(n\cross n)$ identity matrix $\mathbb{I}_{n}$. Similarly, the correlation-function scaling must be \textit{algebraically-decaying}.
\end{enumerate}

In our tensor decomposition solution, tensors $Y$ and $R$ are restricted to being \textit{doubly-unitary}; that is, they are unitary with respect to both vertical and horizontal contractions of each tensor and their Hermitian adjoint. $Q$, on the other hand, is restricted to being only unitary in nature; these relations can be explicitly written in the form 
\begin{align}
\sum_{cd}Y_{acbd}{Y}^{\dagger}_{cdgh} &= \delta_{ag}\delta_{bh} \\ 
\intertext{for the \textit{vertically-contracted} $Y$ tensors,}
\sum_{bd}Y_{acbd}{Y}^{\dagger}_{bdeh} &= \delta_{ae}\delta_{ch}, \\
\intertext{for the horizontally-contracted $Y$ tensors,}
\sum_{kl}R_{iklj}{R}^{\dagger}_{kpql} &= \delta_{ip}\delta_{jq} \\ 
\intertext{for the vertically-contracted $R$ tensors,}
\sum_{jl}R_{iklj}{R}^{\dagger}_{jlqn} &= \delta_{in}\delta_{kq} \\ 
\intertext{for our horizontally-contracted $R$ tensors, and}
\sum_{su}Q_{rtus}{Q}^{\dagger}_{suvw} &= \delta_{rw}\delta_{tv},
\end{align}
for the only possible unitary contraction of the $Q$ tensors. These tensor-diagrammatic relations can be visualized as follows in Figure \ref{multilinear_hyMERA} . In addition, we demand that any reduced-density matrix that is defined in the bulk of a hyMERA network to have a nontrivial entanglement spectrum; in other words, the eigenvalues of a reduced-density matrix must not be proportional to the identity matrix $\mathbb{I}_{n}$ for some causal cone $\mathcal{C}(\mathcal{R}_{n})$. This extra condition ensures the polynomial-like decay of correlations functions in a hyMERA network parameterization; more information on hyMERA can be found in \cite{hyper_evenbly}. 

\begin{figure}
\centering
\includegraphics[width=7.5cm]{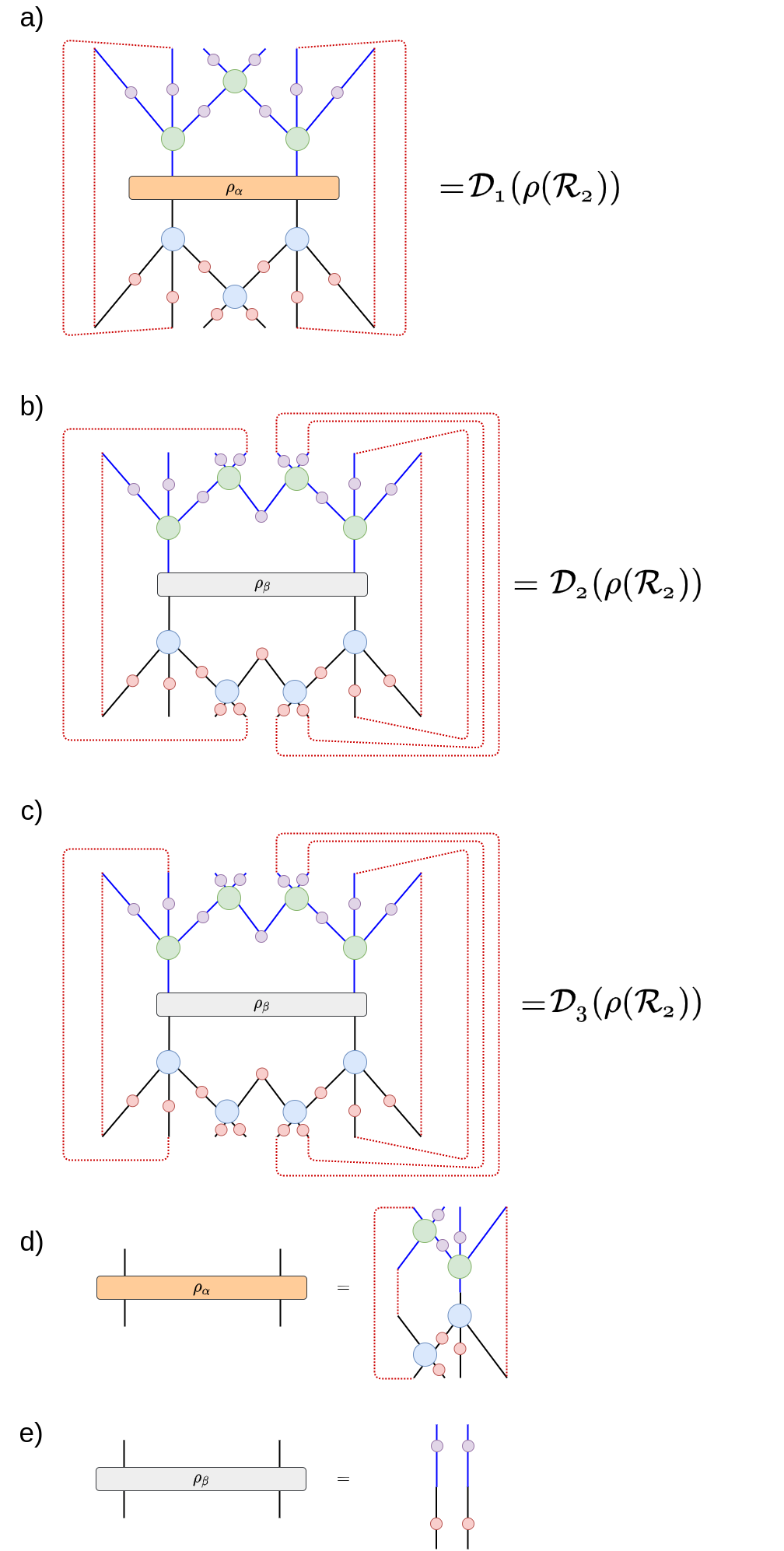}
\caption{Contraction of the descending superoperators in hyMERA are not as computationally costly as in MERA, due in large part to the \textit{holographic causal cone} structure of hyMERA. More details can be found in \cite{hyper_evenbly}. Some of the descending superoperators in the $\{5,4\}$-version of hyMERA are shown above.}
\label{fig:some-descend}
\end{figure}

Contraction and calculation of scaling superoperators in hyMERA is simpler than in MERA, due mainly in part to the unique \textit{holographic causal cone} properties of the network \cite{hyper_evenbly}. Some of the descending superoperators in the $\{5,4\}$-version of hyMERA have been shown in Figure \ref{fig:some-descend}.

Many of the standard procedures applied to the study of MERA-style tensor networks cannot be used currently in the context of a hyMERA network; the reasons for this difficulty lie in three different and related complications: $1)$ a suitable boundary lattice (with a Hamiltonian comprised of localized, nearest-neighbor terms, as in MERA) that can be fed into the scale-invariant portion of the hyMERA network has not been introduced, and $2)$ no efficient methods currently exist for relating a given tensor decomposition (in given $\{p,q\}$ parameterization) to the primaries and descendants of an RCFT.

\section{Quasiperiodic Boundaries and Variational Optimization}\label{challenges_hyMERA}

The Poincar{\'e} disk model is a 2D model, inscribed in the hyperbolic plane $\mathbb{H}^{2}$, in which all points inside the boundary are inside the unit disk and circular arcs inside of the plane lead to orthogonal intersections at the boundary \cite{hyperbolic_text}. Like with the Euclidean plane, it is possible to embed regular polygons into the Poincar{\'e} disk model, covering all of the 2D hyperbolic plane. As noted in Section \ref{hyper_explain} , a Poincar{\'e} disk can be parameterized with Schl{\"a}fli symbols $\{p,q\}$ and fulfills the relation $\frac{1}{p} + \frac{1}{q} < \frac{1}{2}$. As such, a tessellated Poincar{\'e}-disk model exhibits several differences from the non-tessellated version; perhaps the most interesting of which is that the boundary of such a tessellated space does not exhibit periodic boundary conditions \cite{CQCs} , provided that the tilings themselves are regular. The Poincar{\'e} disk model also features the \textit{M{\"o}bius transformations}, which are formed from the M{\"o}bius group $PSL(2,\mathbb{C})$ and leave the manifold invariant.

In the AdS$_{3}$/CFT$_{2}$ correspondence with global AdS coordinates, a timeslice of AdS space can be projected onto a Poincar{\'e} disk, with metric \cite{nastase1}: 
\begin{equation}
ds^{2} = 4\alpha^{2}\frac{d\rho^{2}+\rho^{2}d\phi^{2}}{(1-\rho^{2})^{2}}.
\end{equation}
Here, $\alpha$ is the AdS radius, and the coordinates on the Poincar{\'e} disk take the form $\rho \in [0,1]$. $\phi$ is an angle that is related to the transformations under which a field theory defined on the disk is invariant. A transformation of complex coordinate $z = e^{i\phi}$ of the form $z \mapsto z' = e^{i\theta}\frac{z+w}{w^{*}z+1}$ effectively defines the \textit{M{\"o}bius transformations}. This subgroup of the M{\"o}bius group finds utility in defining the \textit{global conformal transformations} for 2D CFTs \cite{qualls,francesco,conformal_boot1,conformal_boot2,conformal_boot3}, restricted to a timeslice.

Just as in the original Poincar{\'e} disk model, one can tessellate a timeslice of AdS$_{3}$ spacetime; the resulting fractal pattern in the bulk of AdS$_{3}$ breaks the continuous symmetries of the global conformal group \cite{majorana_dimers,central_dimers,qCFT,jahn2021holographictopical} on the boundary. In \cite{holography_tessellate}, it was shown that the holographic dictionary survives the truncation associated with approximating a continuous Poincar{\'e} disk model of AdS$_{3}$ with a regular tessellation; moreover, as described in \cite{CQCs}, a tensor-network realization that imitates a tessellated Poincar{\'e} disk can be described using tensors at every vertex of the bulk. Conversely, edge tensors can be added to a tensor-network manifold that mimics the tessellated Poincar{\'e} disk, as in hyMERA.

In the paradigm of scale-invariant MERA-family tensor networks, preservation of translational invariance is a useful simplification that can aid in the creation of computationally-scalable numerical optimization algorithms in tensor networks, as in general, the cost of representing a local operator grows exponentially with the size of its support \cite{vidal1,evenbly_disordered}. The use of translation-invariant Hamiltonians is exemplified in a MERA-style tensor network, where a Hamiltonian is usually of the form 
\begin{equation}
H = \sum_{i}h(i,i+1), 
\end{equation}
where $h(i,i+1) = h$ is taken to be a nearest-neighbor, local term \cite{vidal2}; such a Hamiltonian exhibit periodic boundary conditions, and an \textit{effective} Hamiltonian $H'$ of the same general form should be located in the same localized region after the application of a realspace RG step \cite{evenbly1}. The expectation value $\bra{\psi}H\ket{\psi}$ is then calculated by numerically descending a reduced-density matrix $\rho$ down through the MERA network, with sequential optimization of the tensor environment within the causal cone \cite{evenbly1,evenbly2,evenbly3}. Please note here that although the Hamiltonian defined on a layer of the scale-invariant MERA network may be comprised of only localized nearest-neighbor terms as in the form above (and is hence translation invariant), this does not imply that the actual MERA network exhibits \textit{exact} translation-invariant symmetry as well. Indeed, it is still an open problem for MERA tensor networks as to whether or not exact translational invariance can be reproduced \cite{evenbly2}. However, it is known that scale-invariant MERA tensor networks exhibit \textit{near} translational invariance, in that they approximately preserve the symmetry. As hyMERA exhibits a highly symmetric bulk and a quasiperiodic structure on the boundary, an appropriately optimized hyMERA would yield emergent translation invariance, but only for a disordered critical Hamiltonian, and not for a \textit{periodic} critical spin chain, like in the case of MERA.

\begin{figure}
\centering
\includegraphics[width=\columnwidth]{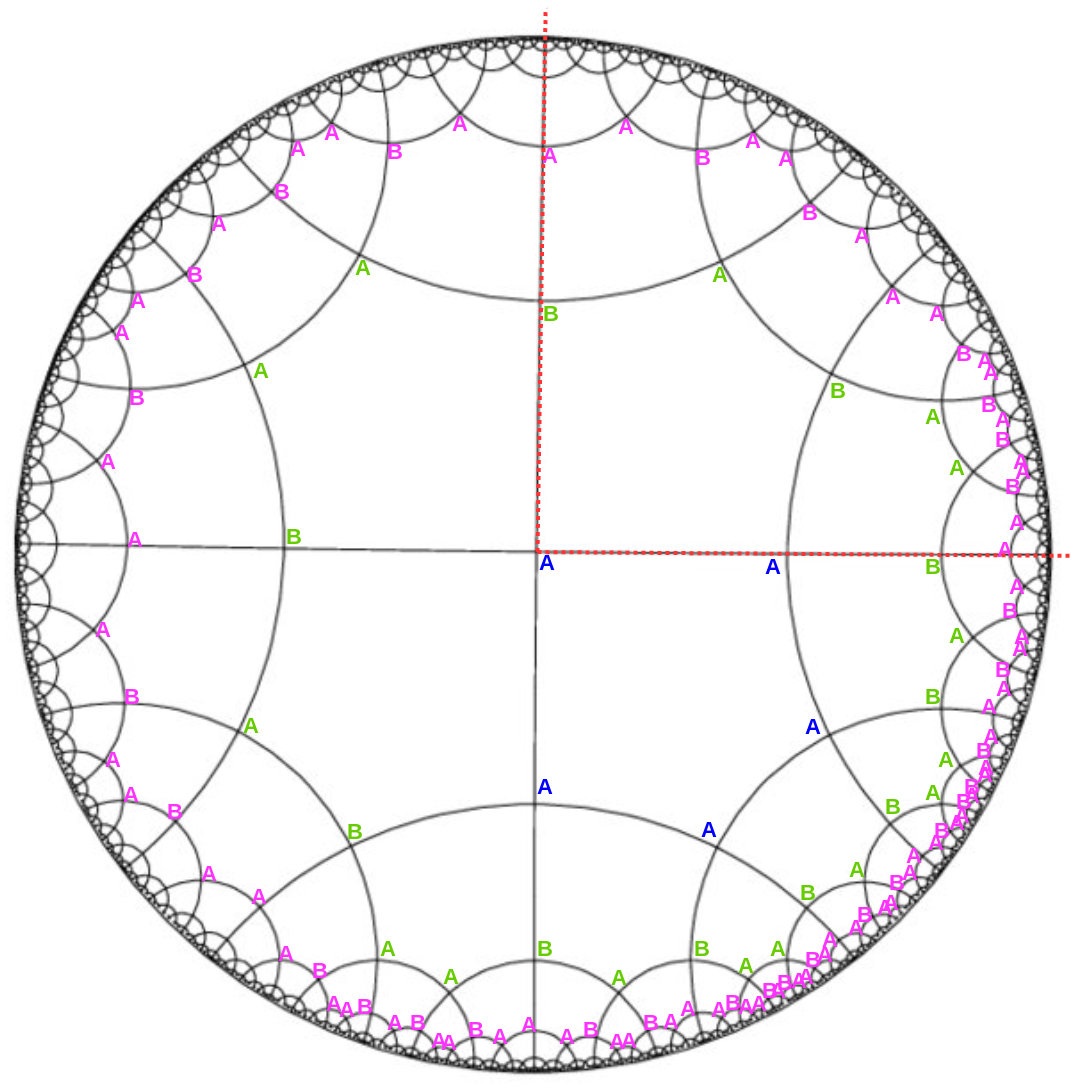}
\caption{A vertex inflation of a $\{5,4\}$ hyperbolic bulk tessellation, which shows how to "grow" a regular tessellation on a hyperbolic manifold in 2D. Blue $A$s represent the first layer of vertex inflation, as shown in the starting pentagon's vertices. Subsequent layers are tiled around the central polygon by applying the necessary inflation rules, as depicted in the green- and magenta-lettered layers of the resulting bulk tessellation. This procedure is described in more detail by \cite{qCFT,CQCs,majorana_dimers,central_dimers,jahn2021holographictopical}. We add tensors to the vertices and edges to the entire bullk of the resultant hyMERA network. If we trace the deflation steps from the boundary toward the center, a string sequence of the form $\{ABABAAB\dots\}$ is observed to renormalize into effective couplings $\{B'A'\dots\}$ as we scale up the network. This effect can be verified by inscribing a causal cone, as the red-dotted lines in the upper-right quadrant show. Subfigures a)-d) show the successive tiling of the Poincar{\'e} disk by applying the inflation rules in a $\{5,4\}$ regular hyperbolic tiling; Subfigure a) begins with the first layering, and subsequent layerings are added as the thermodynamic limit is reached.}
\label{vertex_inflation}
\end{figure}

In a hyMERA, the regular tiling of the Poincar{\'e} disk manifold enforces the loss of translational invariance and periodicity on the boundary of the network. One may still be able to implement variational optimization using a scheme such as the one that we describe below, which uses \textit{conformal quasicrystals} to represent scale steps in the realspace RG transformation. Starting from one of the central pentagons in the $\{5,4\}$ version of hyMERA (shown in Figure \ref{vertex_inflation}), we utilize the \textit{vertex inflation} technique introduced in \cite{CQCs,central_dimers,qCFT,jahn2021holographictopical} to iteratively "grow" a regular tessellation on the Poincar{\'e} disk. The inflation rules for a $\{5,4\}$-regular tessellation are 
\begin{align}
\begin{cases}
a \mapsto abaab, \\
b \mapsto ab.
\end{cases}
\end{align}
We begin by labeling the lower-right central pentagon, and label all of the vertices "$A$", which possess four neighbors each; these are shown in blue-lettered vertices in Figure \ref{vertex_inflation} . The inflation rules are subsequently applied to the following layer surrounding the original pentagon; the vertices are labeled accordingly, as depicted by the green layer of vertex inflation. Finally, the magenta-lettered vertices are assigned letters; this procedure can be carried on until a network of the desired size is constructed. 

Let us now examine the calculation of an expectation value of the Hamiltonian shown below; namely, an expectation value of a Hamiltonian composed of only local, nearest-neighbor terms. Select a string sequence of vertices for a Hamiltonian of the form 
\begin{equation}
H = \sum_{i}J_{i}h(i,i+1), 
\end{equation}
where $J_{i}$ cycles through different coupling constants $J_{b} > J_{a} > 0$, depending on the vertex designation; the constants themselves can vary widely in value, as is expected for strongly-disordered systems \cite{zimboras1,zimboras2}. This Hamiltonian is defined for some layer $\mathcal{L}'$. We can define an effective local Hamiltonian as a substring for a layer $\mathcal{L}$ of the network; such a Hamiltonian is composed of a finite sequence of nearest-neighbor pairs, as in the upper-righthand portion of Figure \ref{vertex_inflation} . Upon application of an RG step, the new effective local Hamiltonian that we find inside the causal cone will be renormalized to new \textit{effective couplings} $\{J_{a'},J_{b'}\}$, which correspond to effective lattice points $A',B'$, respectively. The new effective Hamiltonian would be written down as 
\begin{equation}
H' = \sum_{i}J_{i'}h'(i',i'+1), 
\end{equation}
where $i'$ represents the distance between two nearest neighbors on the layer $\mathcal{L}'$. In this way, it may be possible to implement a realspace RG flow similar to the one in typical MERA-style tensor networks as 
\begin{equation}
\mathcal{H}_{\text{eff}}(a,b) \mapsto \mathcal{H}'_{\text{eff}}(a',b') \mapsto \mathcal{H}''_{\text{eff}}(a'',b'') \mapsto \dots ,
\end{equation}
where the boundary-layer Hamiltonian $\mathcal{H}_{\text{eff}}(a,b)$ is renormalized into the sequential \textit{effective} lattice with a new Hamiltonian $\mathcal{H}'_{\text{eff}}(a',b')$; as described above, the tricky part for such an implementation would be to correctly match renormalization steps to the vertex-inflation rules in the RG step sequence. One may be able to directly implement such a strategy similar to the general outline above, as in \cite{evenbly1,evenbly3}; however, it is unclear whether further complications may arise. These issues are not insurmountable, as it is in general easier to compute expectation values in hyMERA than in a standard MERA tensor network (this simplification is due to the multitensor constraints); however, specific methods have not yet been explored in detail for the context of hierarchical tensor networks built from regular tessellations of bulk hyperbolic manifolds. In addition to the complication described above, there are a few more issues with local Hamiltonian-based variational approaches in hyMERA that also require attention. 

\begin{figure}[ht]
\centering
\includegraphics[width=\columnwidth]{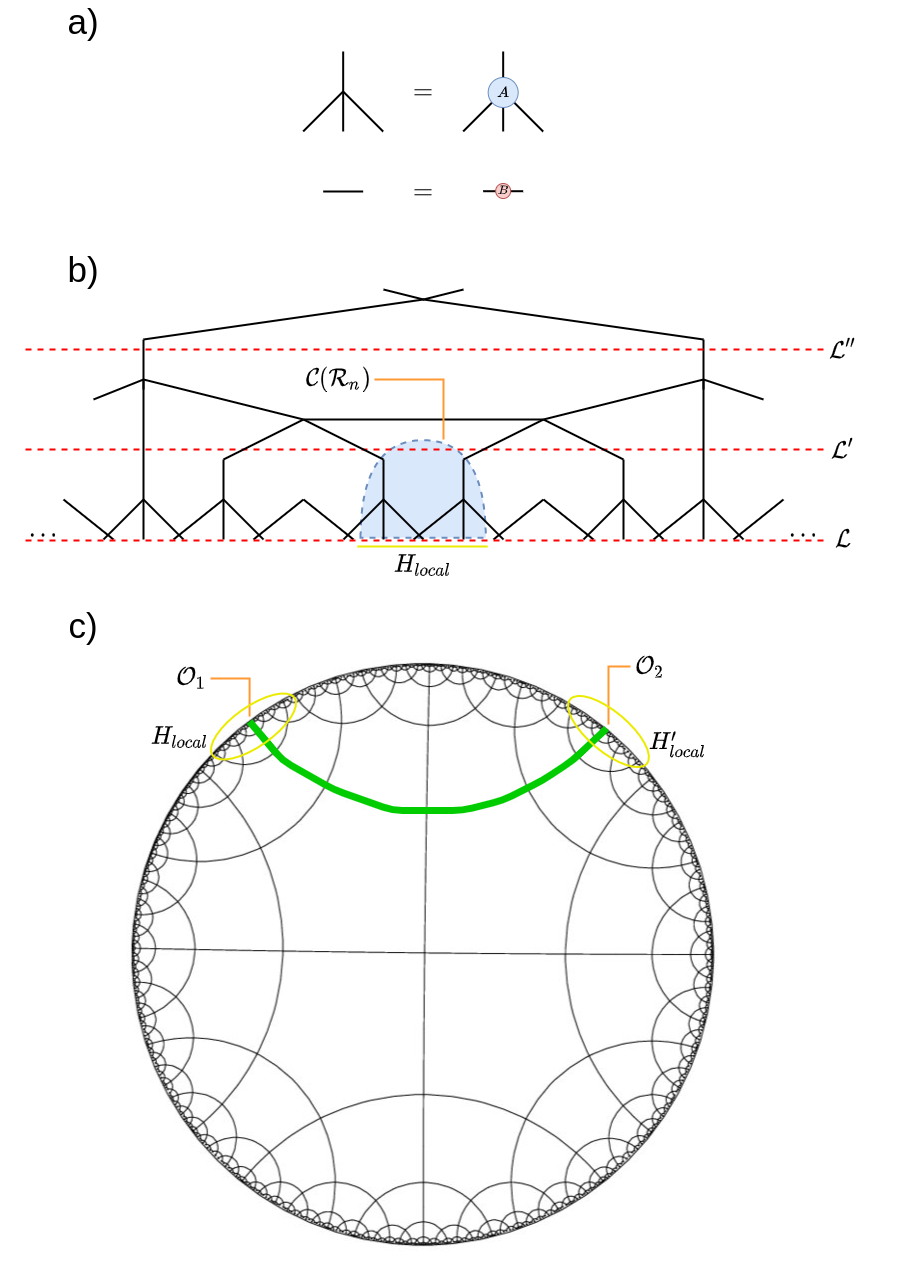}
\caption{Subfigure $b)$ displays the causal-cone setup that one would use in order to minimize for the groundstate expectation value of a local Hamiltonian $H_{local}$; tensors inside of the causal cone can be iteratively optimized by applying the vertex-inflation strategy as described above. However, the same sequence of coupling constants that appear in the local Hamiltonian used for the iterative-optimization procedure is not guaranteed to be found on another subregion of the hyMERA boundary; therefore, one can expect widely varying subsequences of couplings constants (and thus varying local Hamiltonians). As shown in Subfigure $c)$, One may attribute the optimized portion of the network with containing a local Hamiltonian $H_{local}$, as in Subfigure $b)$. As such, $n$-point function calculations may not yield the expected results pertaining to the localized groundstate optimization, since at least one of the observables in the calculation may be located in a subsection of the boundary with a non-matching local Hamiltonian $H'_{local}$. This different local Hamiltonian is comprised of a different subsequence of coupling constants, and thus may not exhibit the same groundstate characteristics as the original boundary subregion.}
\label{causal_opt}
\end{figure}

Firstly, renormalizing a lattice with an appropriate Hamiltonian defined on it may be complicated to implement in a hyMERA network. A typical scale-invariant MERA tensor network requires in practice several \textit{transitional layers} before incorporating the renormalized lattice into the scale-invariant portion of the network. These layers consist of a small set $M$ of $\{(U_{0},W_{0})\dots (U_{M-1},W_{M-1})\}$ tensors which are chosen to be different, in a unique and independent characterization of each layer \cite{evenbly2}. The purpose of these layers is twofold: first, we wish to decouple the bond dimension $\chi$ of the network from the local Hilbert-space dimension $d$ of sites related to the original lattice on which our Hamiltonian is defined; furthermore, we wish to minimize the effect of irrelevant operators in the RG flow once the original lattice has been renormalized into the scale-invariant version of the MERA network. An important question arises, though: given the restricted analytical forms of the tensors in a hyMERA network, how can one devise non-scale-invariant layers of varying bond dimension, in order to study the relevant RG operators in the holographic bulk? This initial step is beneficial in MERA for describing a quantum system by using \textit{only} the relevant RG operators from its pertinent universality class, and it is unclear how such a procedure would be realized in a hyMERA tensor network. 

Secondly, although it may be possible to realize a variational-optimization method in the same vein as in \cite{evenbly2,evenbly3}, it is known that local subsections of quasiperiodic lattices do not encapsulate the global properties of the entire system. This observation further complicates the case for utilizing a local Hamiltonian-based variational scheme, as only the tensors related to the localized boundary region and the causal cone of the hyMERA tensor network will be directly optimized, and the groundstate minimization obtained may not pertain to the entire lattice. We subsequently do not expect non-local quantities such as $n$-point correlation functions to yield the expected results, since at least one of the observables in the $n$-point function would lie on a subregion of the boundary layer where a differing subsequence of coupling constants is present. For example, consider the problem of calculating a $2$-point correlation function in hyMERA after optimizing for some local Hamiltonian $H_{\text{local}}$; performing a localized groundstate optimization as discussed before will not ensure that other portions of the boundary will be optimized for the groundstate of a similar Hamiltonian. In fact, if we try to calculate the $2$-point function between two non-local points which are associated to local Hamiltonians $H_{\text{local}}$ and $H'_{\text{local}}$, we find immediately that it is not clear whether or not the calculated $2$-point function results will be associated with either Hamiltonian, as the localized optimization did not encompass a sufficiently large area for the correlator in question. If we attempt to compensate for this by increasing the size the local Hamiltonian, the computational complexity of optimizing such a region will increase exponentially with the bond dimension \cite{evenbly1,evenbly2,evenbly3}. In any case, such a calculation is irrelevant, because two critical local Hamiltonians with differing quasiperiodic coupling constants cannot share the same $n$-point correlation structure, as this would imply that the same CFT would arise in the continuum limit of both models \cite{henkel}. A diagram of this issue is shown in Figure \ref{causal_opt}. A solution may be possible by incorporating techniques from the \textit{strongly disordered renormalization group} literature \cite{SDRG1,SDRG2}, although this will need to be investigated in more detail.

\section{New Analytical Tensor Forms \& Bounds on CFT spectra}\label{results_section}

In \cite{hyper_evenbly}, only one possible example set of tensors was utilized in order to present hyMERA. In reality, many such solutions can be obtained, demonstrating that solutions to the multitensor constraints themselves are not unique. Two examples of distinct tensor decompositions are shown in Equations \ref{possible-decomp} and \ref{possible-2} ; the usual forms of $\{Q,Y\}$ are kept as those shown in Section \ref{hyper_explain}, but the tensors $T$ and $S$ now are antisymmetric (the original $R$ were symmetric and doubly-unitary), and lack free-parameters along the diagonal and anti-diagonal. The previous set $\{Y,Q,R\}$ contains tensors which are also each endowed with $\mathbb{Z}_{2}$ symmetry, whereas tensors $T$ and $S$ break with this trend. Tensor decomposition sets $\{Y,Q,T\}$ or $\{Y,Q,S\}$ adhere to the same set of multitensor constraints as described before (up to a normalizable constant); moreover, the resulting correlation functions, entanglement spectra of reduced-density matrices, and scaling dimensions of the descending superoperators yield nontrivial relations for both the $\{7,3\}$ and $\{5,4\}$ variants \cite{hyper_evenbly}, in agreement with the constraints needed. These tensor decompositions, as well as many others which were found, are depicted in Figure \ref{fig:generalize}. In addition, more free parameters have been included (i.e., $T(\theta_{6},\theta_{7})$, $S(\theta_{8},\theta_{9})$, whereas $R(\theta_{2})$ in Section \ref{hyper_explain}), showing that new tensor decompositions can be formed which are not necessarily linear transformations of tensors from the original tensor set $\{Y,Q,R\}$. This change brings the total number of free parameters in the $\{7,3\}$ version of hyMERA with sets $\{Y,Q,T\}$, $\{Y,Q,S\}$ to six parameters, and five in the $\{5,4\}$ variant.
\begin{align}\label{possible-decomp}
T_{efgh} &=
\begin{bmatrix}
0 & -\tan^{-1} (\theta_{6}) & -\cos (\theta_{7}) & 0 \\
\tan^{-1} (\theta_{6}) & 0 & 0 & -\cos (\theta_{7}) \\
\cos (\theta_{7}) & 0 & 0 & \tan^{-1} (\theta_{6}) \\
0 & \cos (\theta_{7}) & -\tan^{-1} (\theta_{6}) & 0 \\
\end{bmatrix},
\end{align}
\begin{align}\label{possible-2}
S_{efgh} &= 
\begin{bmatrix}
0 & -\text{cosh} (\theta_{8}) & -\text{cosh} (\theta_{9}) & 0 \\
\text{cosh} (-\theta_{8}) & 0 & 0 & -\text{cosh} (\theta_{9}) \\
\text{cosh} (-\theta_{9}) & 0 & 0 & \text{cosh} (-\theta_{8}) \\
0 & \text{cosh} (-\theta_{9}) & -\text{cosh} (\theta_{8}) & 0 \\
\end{bmatrix}.
\end{align}
\begin{figure*}
\centering
\includegraphics[width=18cm]{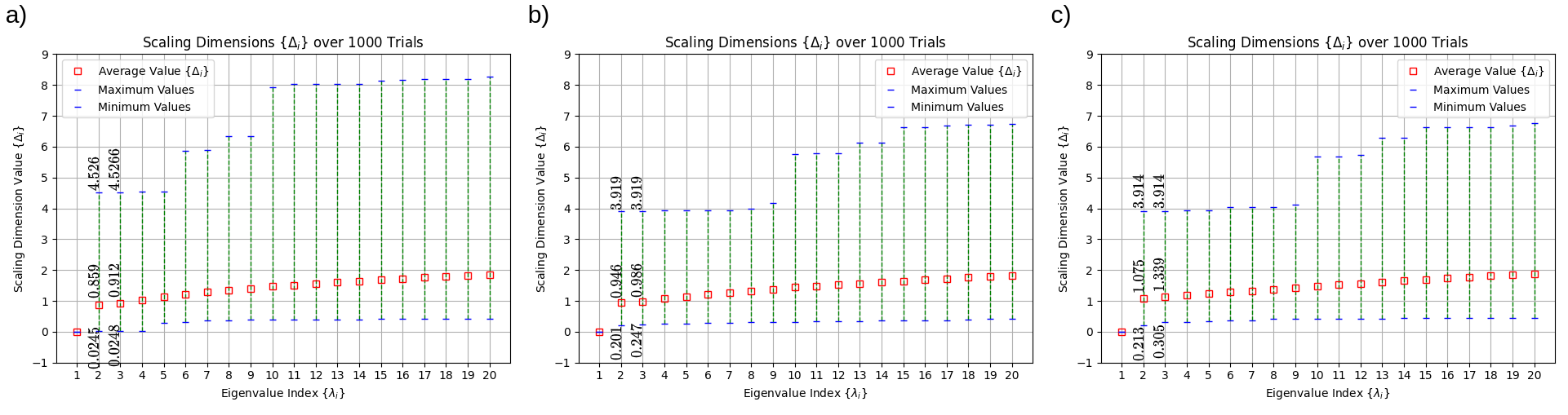}
\caption{Scaling-dimension spectra $\{\Delta_{i}\}$ obtained after 1000 trial runs of diagonalizing the average two-site descending superoperator $\bar{\mathcal{D}}(\mathcal{R}_{2})$ in the $\{5,4\}$-parameterization. We apply the relation $\Delta_{i} = -\log_{s}(\lambda_{i})$ for the tensor decompositions $\{Y,Q,R\}$, $\{Y,Q,T\}$, and $\{Y,Q,S\}$, as shown in Subfigures a), b), and c), respectively. These decompositions are described in Section \ref{hyper_explain}  and \cite{hyper_evenbly}. The results shown suggest that the multitensor constraints themselves do not seem to limit the space of possible scaling dimensions to such an extent that a minimal model CFT would not be simulable, and that proper selection of the tensors in the tensor decomposition set may yield the commensurate scaling dimensions.}
\label{tensor_decompo_show}
\end{figure*}

Figure \ref{tensor_decompo_show}  shows the spectra obtained for all three of the tensor decompositions of $\{5,4\}$-hyMERA after 1000 trials of randomized diagonalizations for the two-site descending superoperator from Subfigure a) in Figure \ref{fig:some-descend}. We subsequently extracted the spectrum of scaling dimensions $\Delta_{i}$. By iterating random values of the free parameters $\{\theta_{i}\}$, one can approximately sample the space of scaling dimensions $\Delta_{i}$ as a function of the eigenvalues $\lambda_{i}$ for the two-site descending superoperator $\mathcal{D}(\mathcal{R}_{2})$. The sampling was employed for the original tensor decomposition in \cite{hyper_evenbly} (wherein the set $\{Y,Q,R\}$ is utilized). Minima and maxima for the first two nontrivial scaling dimensions in Subfigures a), b), and c) in Figure \ref{tensor_decompo_show} demonstrate that the parameter space of possible scaling dimensions can be altered, depending on the decomposition set utilized in the multitensor constraints.
 
\begin{figure}
\centering
\includegraphics[width=\columnwidth]{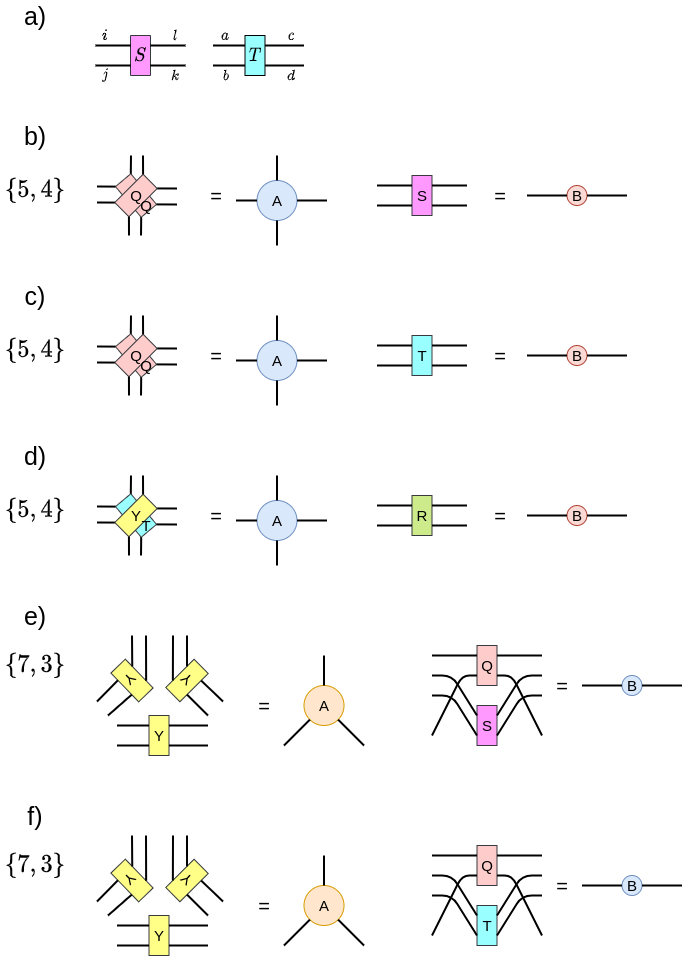}
\caption{a) shows a diagrammatic representation of the two newly-introduced tensors, $S$ and $T$. In b)-d), several variants of hyperinvariant tensor networks are displayed using these antisymmetric variants; in particular, variants b) and c) were subjected to numerical studies, as shown in Figure \ref{tensor_decompo_show}. These variants show that construction of a hyMERA tensor decomposition need not even be composed of strictly unitary tensors, as was previously thought in \cite{hyper_evenbly}.}
\label{fig:generalize}
\end{figure} 
 
Analysis of the average scaling dimensions as a function of their eigenvalues for both tensor decomposition sets shows the scaling dimension values tend to level off, on average. the tensor decompositions $\{Y,Q,R\}$, $\{Y,Q,T\}$, and $\{Y,Q,S\}$ are shown diagrammatically in Figure \ref{fig:generalize}. Our results from individual trials show a considerably larger difference between scaling dimensions, similar to the spectra reported in \cite{hyper_evenbly}. We note that, although it may not be possible to simultaneously extract all of the conformal data using the current tensor decompositions, the maximal and minimal values encapsulate the known primary values for the Ising CFT, as well as several other minimal model RCFTs \cite{henkel}. This observation suggests that it is \textit{in principle} possible that a generalized tensor decomposition exists which may be able to be numerically optimized for an RCFT's conformal data. Furthermore, we have shown that imposing the multitensor constraints themselves does not appear to limit the space of possible scaling-dimension values to the exclusion of certain minimal model CFTs, as was previously surmised \cite{hyper_evenbly}.

There are several open questions related to tensor decompositions in a hyMERA network: a) what types of limitations exist for a given tensor decomposition $\{\mathcal{T}_{1}\dots \mathcal{T}_{n}\} \in \{A,B\}$; b) how to generalize the multitensor constraints; c) what characteristics of tensor decompositions are amenable for simulating the entanglement properties of CFTs; and d) how to characterize the quality of CFT simulations as a function of some tensor decomposition $\{A,B\}$, in comparison to another decomposition $\{A',B'\}$. Answering these questions may help to understand better the relationship between the distinct physical properties of a CFT and a tensor network that simulates it.

\section{Discussion}\label{discussion-section}

In this work, we have presented suitable critera for both a tensor generalization and variational optimization in hyMERA. We have accomplished this analysis as follows: Firstly, we have examined the standard Hamiltonian-based variational algorithms from MERA \cite{evenbly1,evenbly2,evenbly3}, and have shown that modifications are required in order to account for the boundary quasiperiodicity inherent to regular tessellations of hyperbolic bulk manifolds. Additionally, we have addressed the issues associated with transitional layers in hyMERA, and have provided possible solutions. Secondly, we have provided two new analytical forms of tensors, $S$ and $T$, which adhere to the multitensor constraints, even without sharing many of the features of the original tensors $Y,Q,$ and $R$. We have demonstrated that solutions to the multitensor constraints are neither unique (as was suggested in \cite{hyper_evenbly}), nor difficult to find, emphasizing that a generalization of the strict analytical forms used up until now in hyMERA likely exists. Lastly, we have performed a randomized analysis of one of the descending superoperators in hyMERA with various tensor decompositions, and have shown that the bounds on the scaling-dimension spectra change as we vary the constituent tensors in a given tensor decomposition. Futhermore, it was found that all of the tensor decompositions analyzed in this work exhibit bounds that encapsulate many of the known minimal model CFTs, such as the Ising, $3$-state Potts, and the tricritical Ising CFT models. These foundational results illustrate that hyperinvariant tensor networks, if generalized properly, could be used as a numerical tool for analyzing CFTs arising from hyperbolic tessellations with quasiperiodic boundaries, rather than simply a unique, paradigmatic example which combines aspects of HaPPY and MERA tensor networks for the purpose of simulating the AdS/CFT correspondence.

Although we have shown evidence that other analytical tensor forms in hyMERA do exist, we do not know what physical interpretation our new proposed tensor sets exhibit in terms of CFT reproducibility, if any. We observed that the space of allowable scaling dimensions does change significantly upon permuting the various tensors into different sets; however, it is unclear how the strict analytical forms of a tensor-decomposition set relate to the spectra of minimal model CFTs. In fact, given a suitable generalization of the analytical forms used in hyMERA, one may be able to understand how to properly reproduce a CFT spectrum in hyMERA. We leave such an investigation for future work.

Of the still many open questions related to understanding hyperinvariant tensor networks, the interpretation of hyMERA as a \textit{holographic quantum error correction code} is certainly intriguing. Some recent work \cite{master_thesis,cao2021hyperinvariant} has suggested associating hyMERA to codes derived from perfect tensor networks, although a more precise formulation is needed in order to properly characterize hyMERA as an error-correcting code. An interesting avenue may be to examine multitensor-constraint generalizations (and implications for quantum error correction in hyMERA) using the tools from recent work connecting Majorana dimer fermions, AME states, and the \textit{holographic pentagon code} (HyPec) \cite{majorana_dimers,central_dimers,happy,jahn2021holographictopical,AME,AME1,AME2,AME3,AME4}.

Finally, one may be able to leverage alternative methods from high-energy physics in order to realize and implement a variational optimization scheme in hyMERA, if a more general description of the analytical tensor forms described can be formulated. One such possibility could be seen in leveraging the \textit{conformal bootstrap} \cite{conformal_boot1,conformal_boot2,conformal_boot3,zamolodchikov}.

\section{Acknowledgements}

We thank Santiago Oviedo-Casado, Javier Molina-Vilaplana, Matthias Volk, Sukhbinder Singh, and Michael Kastoryano for insightful discussions, as well as Alexander Jahn, Jens Eisert and Glen Evenbly for their feedback on the manuscript. J.P. is grateful for financial support from Ministerio de Ciencia, Innovaci{\'o}n y Universidades (SPAIN), including FEDER (Grant Nos. PGC2018-097328-B-100) together with Fundaci{\'o}n S{\'e}neca (Murcia, Spain) (Project No. 19882/GERM/15).

\clearpage

\bibliography{bibliography}

\clearpage

\appendix 
\setcounter{figure}{0}

\section{MERA Tensor Networks}\label{MERA_family}

\begin{figure}[h]
\centering
\includegraphics[width=\columnwidth]{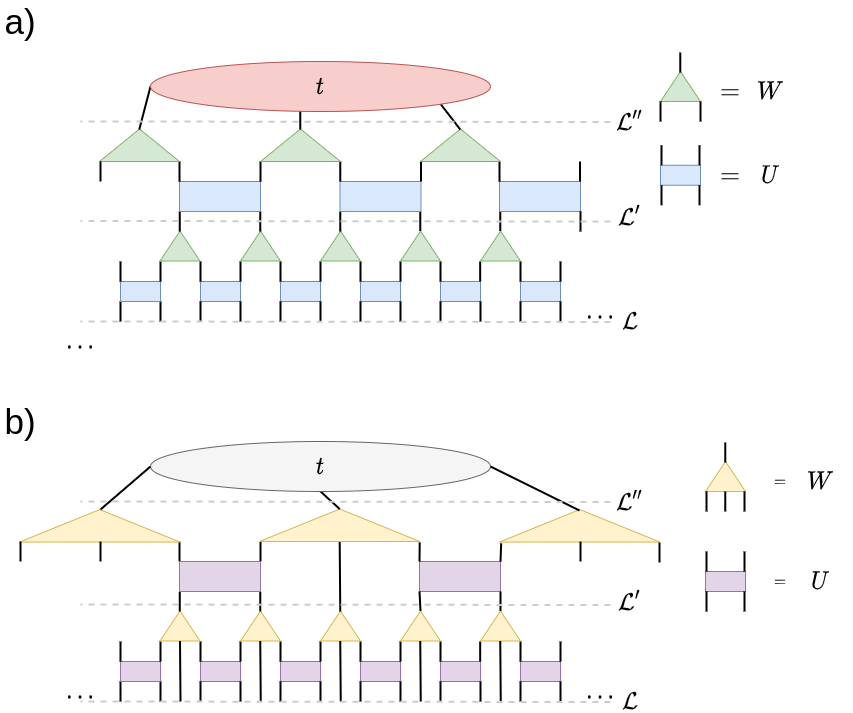}
\caption{Subfigures $a)$ and $b)$ show the binary MERA and the ternary MERA implementations \cite{vidal1,vidal2,evenbly1,evenbly2,evenbly3}. Top tensors are shown as tensors $t$ in the figure, whereas the isometries and unitaries are identified by key.}
\label{MERA_family_TNs}
\end{figure}

The \textit{multiscale entanglement renormalization ansatz} (MERA) has attained much success as a tensor-network ansatz that implements real-space renormalization group techniques in order to facilitate the study of \textit{scale-invariance} and \textit{quantum criticality} in the context of quantum lattice models. Several varieties of MERA-family tensor networks are shown in the Figure \ref{MERA_family_TNs} ; however, for the purposes of this article, we shall focus our attention on \textit{binary} MERA and \textit{ternary} MERA. The basic idea of these tensor networks is the same; a MERA tensor network consists of several types of tensors:
\begin{align}
\intertext{$1)$ Top Tensors:} 
t = U\ket{0}\ket{0}, (t)_{\mu\nu} = (U)^{\alpha\beta}_{\mu\nu}\rvert_{\alpha,\beta = 0}, \\
\intertext{where two indices are contracted, and each top tensor corresponds to a two-body unitary operator. The top tensor is normalized to one;}
\intertext{$2)$ Isometries:}
\sum_{\mu\nu}W_{\alpha}^{\mu\nu}W^{\alpha'}_{\mu\nu} = \delta_{\alpha\alpha^{'}}, 
\intertext{if only $2$ indices are contracted,}
\sum_{\alpha}W_{\alpha}^{\mu\nu}W^{\alpha}_{\mu'\nu'} = \mathcal{P}^{\mu\nu}_{\mu^{'}\nu^{'}},
\intertext{in the case of $1$ index being contracted;}  
\intertext{and $3)$ Disentanglers:} 
\sum\limits_{\mu\nu}(U^{*})_{\alpha\beta}^{\mu\nu}(U)^{\alpha^{'}\beta^{'}}_{\mu\nu} = \sum\limits_{\mu\nu}U_{\alpha'\beta'}^{\mu\nu}(U^{*})^{\alpha\beta}_{\mu\nu} = \delta_{\alpha\alpha^{'}}\delta_{\beta\beta^{'}},
\end{align}
where the disentanglers are simply unitarity gates\cite{vidal1,vidal2}.

These tensors are organized in a hierarchical fashion with the goal of implementing a local, \textit{real-space} renormalization group transformation; in the scale-invariant version of MERA, all of the set of $\{U,W\}$ that comprise the network are defined to be exactly the same. This renormalization group flow can be described as 
\begin{equation}
(\mathcal{L},\hat{H}) \rightarrow (\mathcal{L}',\hat{H}') \rightarrow (\mathcal{L}'',\hat{H}'') \rightarrow \dots,
\end{equation}
where $\mathcal{L}'$ refers to the first effective lattice generated upon promoting an operator (such as the locally separable Hamiltonian $\hat{H}$ in the equation below) to a higher layer, inside of the bulk of the MERA network. 

An important stipulation for realizing the RG flow in MERA comes from the restrictions on the Hamiltonian. In MERA, it is assumed that the Hamiltonian can be broken down into a sum of nearest-neighbor terms: 
\begin{equation}
\hat{H} = \sum_{s}h(s,s+1),
\end{equation}
where $h(s,s+1) = h$ is a localized Hamiltonian term. Every portion of the lattice, then, can be viewed as completely characterizing $\hat{H}$.

Renormalization of a local operator or reduced-density matrix can be realized in either form of MERA via \textit{ascending} and \textit{descending} superoperators, respectively. In a typical non-scale-invariant MERA, the ascending and descending superoperators are responsible for sequentially promoting/demoting an operator up/down the MERA: 
\begin{align*}
\mathcal{O} &\xrightarrow{\mathcal{A}(\mathcal{O})} \mathcal{O}' \xrightarrow{\mathcal{A}(\mathcal{O}')} \mathcal{O}'' \xrightarrow{\mathcal{A}(\mathcal{O}'')} \dots, \\
\intertext{ we see a similar expression for descending superoperators $\mathcal{D}$:}
\dots &\xrightarrow{\mathcal{D}(\rho''')} \rho'' \xrightarrow{\mathcal{D}(\rho'')} \rho' \xrightarrow{\mathcal{D}(\rho')} \rho,
\intertext{ and we can conclude from the action of both of these superoperators that they are in fact dual to each other (in the Choi-Jamiolkowski representation of a quantum channel \cite{nielsen1,PRL-QuantumMERAchannels,wolf1}):}
\mathcal{D} &= \mathcal{A}^{*} \\ 
\intertext{or}
tr\bigg[ \mathcal{O}_{s-1}\mathcal{D}(\rho_{s})\bigg] &= tr\bigg[ \mathcal{A}(\mathcal{O}_{s-1})\rho_{s}\bigg].
\end{align*}
However, for the purposes of this paper, we consider the scale-invariant regime; as such, every step in the renormalization group calculus now maps from fixed RG point to fixed RG point. The form of the descending and ascending superoperators simplifies to that of the \textit{scaling superoperators}, which are described as:
\begin{align}
\mathcal{A}(\mathcal{O}) \mapsto \mathcal{S}(\phi_{\alpha}), \\
\intertext{ which is given by}
\mathcal{S}(\phi_{\alpha}) = \lambda_{\alpha}\phi_{\alpha}, \\
\intertext{where $\phi_{\alpha}$ is a scaling field operator, and $\lambda_{\alpha}$ are the eigenvalues after application of the superoperator. Addtionally, we have that}
\Delta_{\alpha} = -\log_{s}(\lambda_{\alpha}), 
\end{align}
where $\Delta_{\alpha}$ are the scaling dimensions of a CFT, and $s$ is the scale factor, determined by number of sites renormalized per layer. Additionally, the dual of the ascending superoperator $\mathcal{A}^{*}$, the descending superoperator $\mathcal{D}$, is described in the scale-invariant regime as $\mathcal{S}^{*}(\rho(\mathcal{R}_{n}))$. The scaling superoperators of a MERA can be diagonalized in order to estimate the scaling dimensions of a related CFT \cite{evenbly1,evenbly2,evenbly3,PRL-QuantumMERAchannels}. 

The $2$- and $3$-point correlation functions in MERA can be calculated in order to find three-point structure constants, as well; this is done after first performing a variational optimization of the groundstate energy for a particular quantum critical model, such as the transverse-field Ising model \cite{evenbly2,evenbly3}. Afterwards, the two- and three-point functions are seen to have the general forms: 
\begin{equation}
\big< \phi_{\alpha}\phi_{\beta}\big> = \frac{\lambda_{\alpha\beta}}{|r_{xy}|^{\Delta_{\alpha}+\Delta_{\beta}}},
\end{equation}
for the two-point function, and the three-point function as:
\begin{equation}
\big< \phi_{\alpha}\phi_{\beta}\phi_{\gamma}\big> = \frac{\lambda_{\alpha\beta\gamma}}{|r_{xy}|^{\Omega^{\gamma}_{\alpha\beta}}|r_{yz}|^{\Omega^{\alpha}_{\beta\gamma}}|r_{zx}|^{\Omega^{\beta}_{\gamma\alpha}}},
\end{equation}
where $\Omega^{\gamma}_{\alpha\beta} = \Delta_{\alpha} + \Delta_{\beta} - \Delta_{\gamma}$, $\{\phi_{\alpha}\}$ represent scaling operators from the relation $\mathcal{S}(\phi_{\alpha}) = \lambda_{\alpha}\phi_{\alpha}$, and $|r_{xy}| = |x-y|$. Although these general forms were obtained for two- and three-point functions in MERA after groundstate-energy minimization, these relations are completely general for all 2D CFTs \cite{qualls,francesco,henkel,on_the_plane}.

In \cite{evenbly3} it was shown that the \textit{central charge} of a $(1+1)$-dimensional CFT can be calculated using the Cardy formula for entanglement entropy \cite{cardy,kitaev}:
\begin{equation}
\mathcal{S}(\rho(\mathcal{R}_{n})) - \mathcal{S}(\rho(\mathcal{R}_{m})) = \frac{c}{3}\big( \log{n} - \log{m}\big),
\end{equation}
where $n > m$, and $n,m \in \mathbb{Z}_{+}$. $\mathcal{S}(\rho(\mathcal{R}_{n}))$ is the $n$-site entanglement entropy, given by the von Neumann entropy $\mathcal{S}(\rho(\mathcal{R}_{n})) = -Tr\big[\rho(\mathcal{R}_{n})\log(\rho(\mathcal{R}_{n}))\big]$. 

The actual reduced-density matrices in MERA are associated with \textit{causal cones} $\{\mathcal{C}(\mathcal{R}_{n})\}$ that depend on the size of the lattice that we are considering at an $n$-site boundary $\mathcal{R}_{n}$. 

It was Swingle \cite{swingle1,swingle2} who first realized that some features of the AdS/CFT correspondence are similar to what is proposed in MERA; among these include the emergence of a discretized timeslice of AdS space. However, MERA is not generally seen as an \textit{exact} discrete analog of AdS space; work from \cite{beny1} pointed out that MERA actually may be more suitable as a discretized version of \textit{de Sitter} space, rather than anti-de Sitter space. Additionally, Carroll et al. \cite{carroll1} showed that MERA is only capable of describing states at length scales larger than the AdS radius, and, that the inherent directionality of MERA's isometric and unitary tensors provide intractable bounds for reproducing any notion of the correspondence within the current version of MERA. Modifications of MERA, however, were not ruled out as potentially fulfilling some of the necessary requirements in order to accurately reproduce holographic state correspondences \cite{czech_defect,evenbly_minimal,Caputa_2017,Takayanagi_2018,Caputa_2017_2,Miyaji_2017,Nozaki_2012,Miyaji_2015}, as MERA was also provided an interpretation pertaining to \textit{kinematic space discretizations} \cite{czech_kinematic}. Recently, a lightcone interpretation of MERA was introduced as well \cite{Evenbly_2011,milsted2018geometric,milsted2018tensor,milsted2018tensorpath}.

\section{Perfect Tensor Networks}\label{perfect_tensors}

After the proposal of \cite{harlow_bulk}, which stated that many features of holography can be reinterpreted in the language of quantum error correction, Harlow et al.\cite{happy} proposed a new class of tensor network that is based on the construction of a quantum error correction code in the tensor-network framework. The elementary building blocks of such a tensor network are \textit{perfect tensors}, a special class of tensor that exhibits maximal entanglement along any bipartition. More succinctly, perfect tensors are defined as specific types of isometric tensors (which are defined naturally as a map $\mathcal{T}$ from one Hilbert space $\mathcal{H}_{a}$ to another $\mathcal{H}_{b}$ such that $\mathcal{T}:\mathcal{H}_{a} \mapsto \mathcal{H}_{b}$, and $dim(\mathcal{H}_{a}) \leq dim(\mathcal{H}_{b})$), where 
\begin{align}
\mathcal{T}:\ket{a} \mapsto \sum_{b}\ket{b}\mathcal{T}_{ba}, \\
\intertext{where $\{\ket{a}\},\{\ket{b}\}$ are the complete orthonormal bases for $\{\mathcal{H}_{a},\mathcal{H}_{b}\}$, respectively, and}
\sum_{b}T^{\dagger}_{a'b}T_{ab} &= \delta_{aa'}, \\ 
\intertext{and}
\sum_{b}T_{ab}T^{\dagger}_{a'b} &= \mathcal{P},
\end{align}
and $\mathcal{P}$ is a projective operator. A perfect tensor, then, is defined as a tensor with $2n$ indices (where $n$ is the number of qubits), and, for a bipartition of indices into sets:
\begin{align}
\mathbb{A}| \leq |\mathbb{A}^{c}|, \\ 
\intertext{the magnitudes of each set must sum up to the total $z$:}
|\mathbb{A}| + |\mathbb{A}^{c}| &= z, \\ 
\intertext{and $\mathcal{T}$ must still be proportional to an isometric tensor, i.e.}
\sum_{b}\mathcal{T}^{\dagger}_{a'b}\mathcal{T}_{ab} &= C\delta_{aa'}, \\ 
\intertext{ where $C$ is a constant, and}
\sum_{b}\mathcal{T}_{ab}\mathcal{T}^{\dagger}_{a'b} &= C\mathcal{P}.
\end{align}
However, not all $2n$-index states possess a perfect tensor state \cite{AME,AME1,AME2}.

\begin{figure}
\centering
\includegraphics[width=\columnwidth]{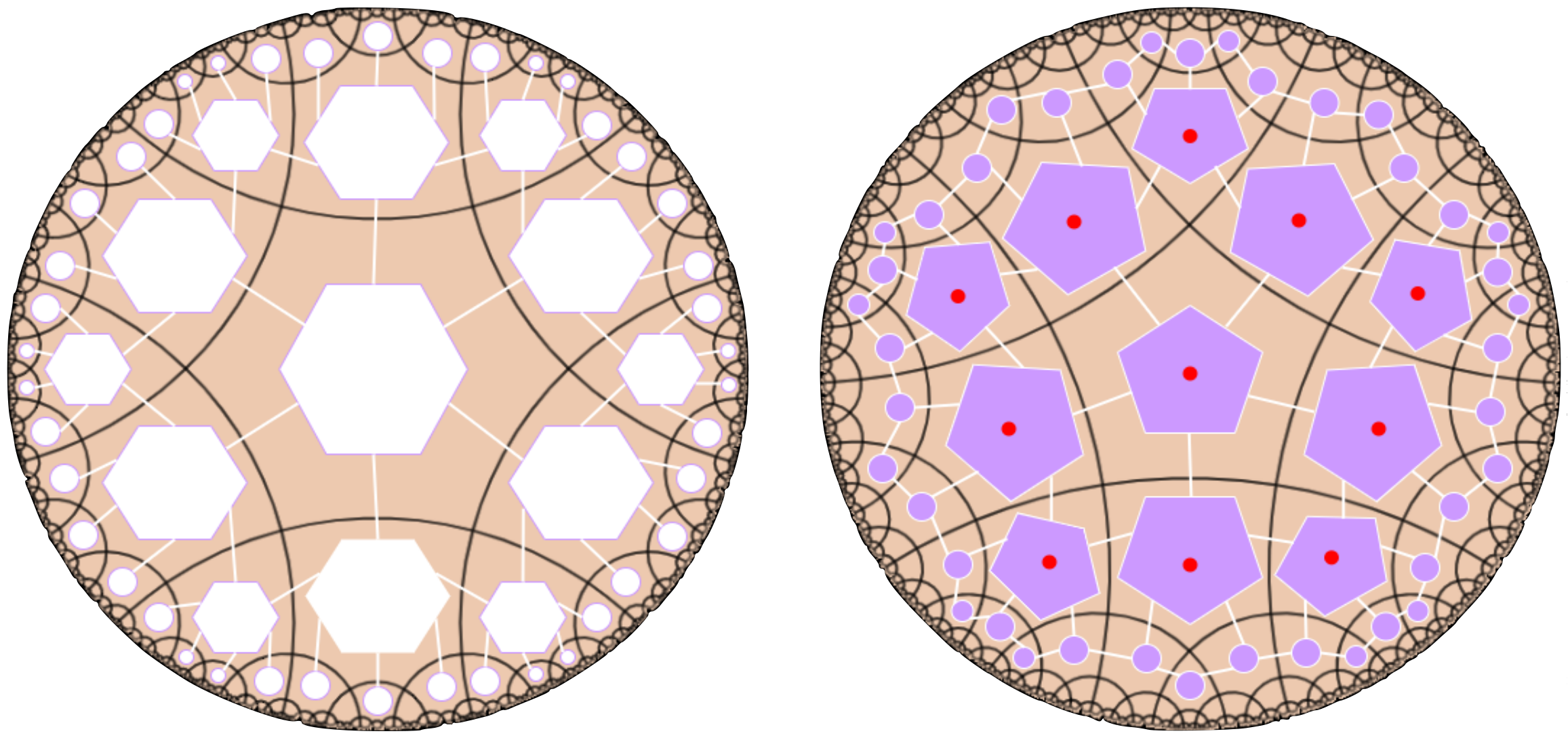}
\caption{Two examples of perfect tensor networks: the hexagonal state, as originally described in \cite{happy}, which describes a \textit{pure holographic state}, and the pentagonal code (also referred to as the \textit{HyPeC} in \cite{majorana_dimers,central_dimers,jahn2021holographictopical,qCFT}), which contains a bulk logical index for encoding a $[[5,1,3]]$ \textit{holographic quantum error correction code}. The HyPeC itself spurred the development of several other related \textit{holographic tensor network codes} \cite{brennen1,farrelly2020tensornetwork,farrelly2020parallel,Harris_2020}. It should be noted here that the Schl{\"a}fli number convention used in the perfect tensor literature follows the scheme $\{q,p\}$, whereas for hyMERA, the convention follows the form $\{p,q\}$. As such, the hexagonal state shown on the left is embedded on a $\{4,6\}$ manifold tessellation; the HyPeC is embedded on a $\{4,5\}$ manifold tessellation.}
\label{tilings}
\end{figure}

The fact that these tensor networks are comprised of perfect tensors gives the bulk portion of the tensor network a high degree of symmetry; this idea is illustrated with the concept of \textit{bulk tensor pushing}. More simply, an operator $\mathcal{O}$ acting on an input leg of an isometric tensor can be represented by an equal-norm operator $\mathcal{O}'$ on the output leg. We see this property by the following:
\begin{align}
\mathcal{T}\mathcal{O} = \mathcal{T}\mathcal{O}\mathcal{T}^{\dagger}\mathcal{T} = (\mathcal{T}\mathcal{O}\mathcal{T}^{\dagger})\mathcal{T} = \mathcal{O}'\mathcal{T}.
\end{align}
One can "push" operators through a network of perfect tensors by exploiting this characteristic. The extra symmetry of the perfect tensors in the bulk provides us with the isotropic symmetry that we seek in an actual timeslice of AdS space. This quality provides a unique advantage over MERA in terms of describing the bulk characteristics of holography. Additionally, the enhanced symmetry in the bulk allows one to define a stabilizer formalism for the purpose of encoding quantum information in the form of a quantum error correction code\cite{happy}.

It has been shown that when perfect tensor networks are assigned to a bulk tessellation, associated with a strongly-disordered boundary in the form of a \textit{conformal quasicrystal} \cite{CQCs}, perfect tensor states can \textit{on average} observe certain characteristics reminiscent of CFT-groundstates \cite{central_dimers,majorana_dimers,qCFT}. It was shown in several works that perfect tensor states can be related to Majorana dimer states \cite{majorana_dimers,central_dimers}, while still retaining properties of exact quantum error correction. Indeed, such Majorana dimer-based models have been associated with critical Hamiltonians in the lattice limit \cite{central_dimers,qCFT}. However, entanglement relationships in these models are typically sparse; an averaging procedure is required in order to observe CFT-like behavior. This work advances the results found in \cite{happy,infinitehappy}, where $n$-site correlation functions previously were shown to result in either zero or a phase, without any averaging treatment. Finally, a holographic analog to the \textit{Bacon-Shor} quantum error correction code was proposed in \cite{cao2021approximate}.

\end{document}